\documentclass[runningheads]{llncs}
\usepackage{graphicx}
\usepackage{xcolor}
\usepackage{array}
\usepackage{amsmath}
\usepackage{pifont}
\usepackage{multirow}
\usepackage{todonotes}
\usepackage{threeparttable}
\usepackage{hyperref}
\usepackage{draftwatermark}

\begin{document}
	
	\SetWatermarkText{PrePrint}
	\SetWatermarkScale{5}

\hrule
This is a preprint version that has not undergone peer review or any post-submission improvement or corrections. The copyright lies with Springer. The final publication is published in Applied Reconfigurable Computing : Architectures, Tools and Applictions (ARC2024) and is available online at Springer at \href{}{https://doi.org/10.1007/978-3-031-55673-9\_4}
\hrule 
\title{Reconfigurable Edge Hardware for Intelligent IDS: Systematic Approach
\thanks{This work has been partially funded by the German Ministry of Education and Research (BMBF) via project RILKOSAN (16KISR010K) and partially via project SILGENTAS (16KIS1837).}}
\author{Wadid Foudhaili* \orcidID{0009-0004-3544-6536} 
\and Anouar Nechi \orcidID{0000-0001-9680-6145} 
\and Celine Thermann \orcidID{0009-0004-2354-1616} 
\and Mohammad Al Johmani \orcidID{0009-0009-9126-0919}
\and Rainer Buchty \orcidID{0009-0004-9413-2078} 
\and Mladen Berekovic \orcidID{0000-0003-1911-756X} 
\and Saleh Mulhem \orcidID{0000-0001-7380-5270}
}
\authorrunning{W. Foudhaili et al.}
\institute{Institute of Computer Engineering(ITI), Universit\"{a}t zu L\"{u}beck, L\"{u}beck, Germany
Corresponding author: \email{wadid.foudhaili@uni-luebeck.de}\\}
{\let\newpage\relax\maketitle}
%\maketitle              

\newcommand{\tdi}[1]{\todo[inline]{#1}}

\begin{abstract}
Intrusion detection systems (IDS) are crucial security measures nowadays to enforce network security. Their task is to detect anomalies in network communication and identify, if not thwart, possibly malicious behavior. Recently, machine learning has been deployed to construct intelligent IDS. This approach, however, is quite challenging 
particularly in distributed, highly dynamic, yet resource-constrained systems like Edge setups. In this paper, we tackle this issue from multiple angles by analyzing the concept of intelligent IDS (I-IDS) while addressing the specific requirements of Edge devices with a special focus on reconfigurability. Then, we introduce a systematic approach to constructing the I-IDS on reconfigurable Edge hardware. For this, we implemented our proposed IDS on state-of-the-art Field Programmable Gate Arrays (FPGAs) technology as (1) a purely FPGA-based dataflow processor (DFP) and (2) a co-designed approach featuring RISC-V soft-core as FPGA-based soft-core processor (SCP). We complete our paper with a comparison of the state of the art (SoA) in this domain. The results show that DFP and SCP are both suitable for Edge applications from hardware resource and energy efficiency perspectives. Our proposed DFP solution clearly outperforms the SoA and demonstrates that required high performance can be achieved without prohibitively high hardware costs. This makes our proposed DFP suitable for Edge-based high-speed applications like modern communication technology.

\end{abstract} 

\section{Introduction}
An Intrusion Detection System (IDS) is crucial in fortifying network security, such as, but not limited to inflicted Distributed Denial-of-Service attacks (DDoS). Basically, an IDS acts as an additional defense layer, detecting and responding to potential threats that may elude preemptive measures. It is also defined \cite{vasilomanolakis2015taxonomy,denning1987intrusion} as a security tool that constantly monitors host or network traffic or both to detect any suspicious behavior that violates the security policy and compromises its confidentiality, integrity, and availability. The typical outcome of the system is to generate alerts about detected malicious behavior to the host or network administrators.

A successful DDoS attack that was reported in 2016 \cite{Forbes2016} leaves us with the following conclusion \emph{``If there was a distributed intrusion detection system, it might have been able to detect the attack at its early stage and limit the loss caused by the attack.''}\cite{sha2020survey}. This capability of distributed IDS encompasses identifying malware, phishing attacks, and other cyber threats in an interactive manner. 

\begin{figure}[!ht]
    \centering
    \includegraphics[width=0.75\linewidth,height=.5\linewidth]{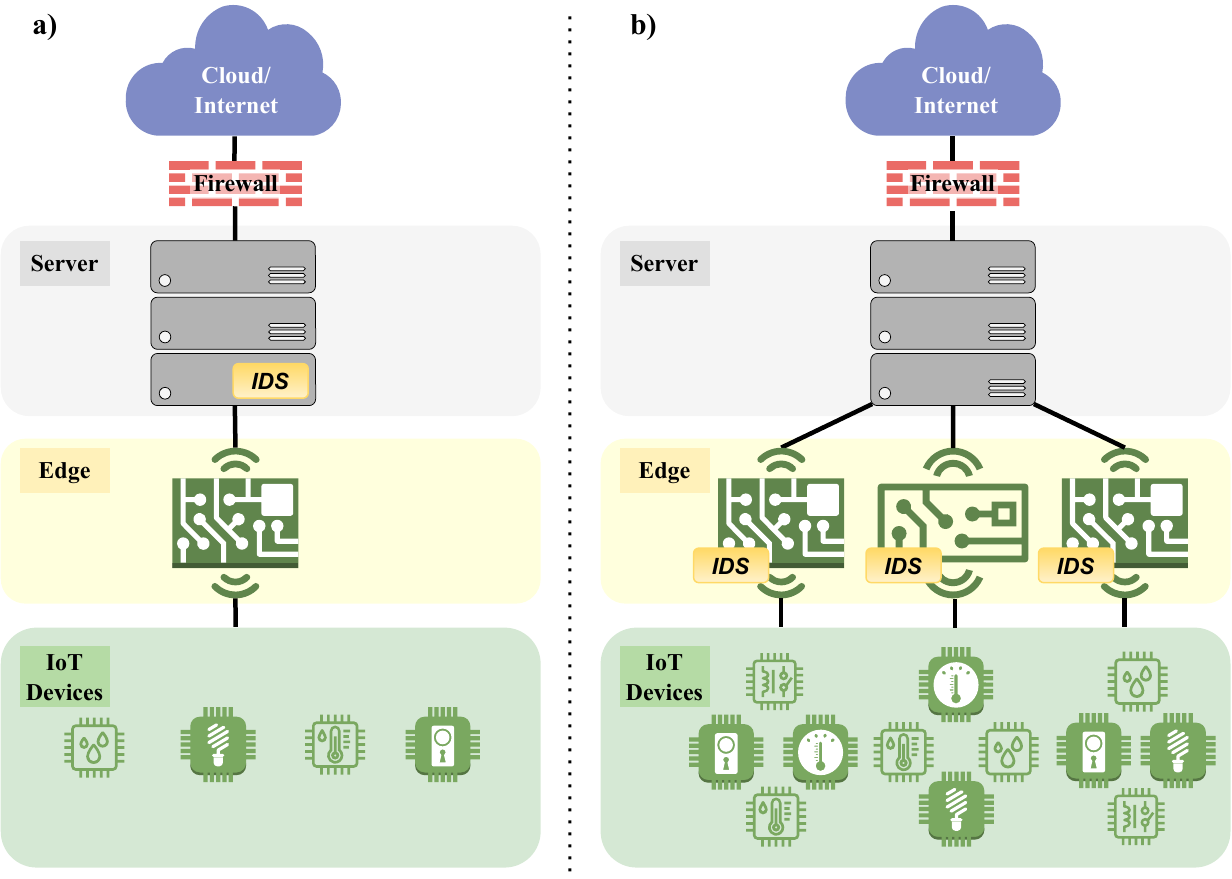}
    \caption{(a) Conventional IDS vs (b) Distributed IDS on the Edge}
    \label{fig:IDS_cats}
\end{figure}

Fig.~\ref{fig:IDS_cats} shows two deployment scenarios of IDS that we call \textit{conventional} compared to \textit{Distributed Edge-based}. The distributed IDS should satisfy special requirements to meet the hardware and power constraints of the edge level. However, it should be noted that conventional IDS leveraging reconfigurable hardware dramatically improves the detection system's performance \cite{1106666}. Therefore, reconfigurable hardware such as field-programmable gate arrays (FPGAs) has become one of the foundations for IDS on the Edge as well.    
\subsection{Machine Learning-based IDS on the Edge}
Machine learning (ML) models, particularly deep neural networks (DNN), have shown a potential to enhance the performance of intrusion detection mechanisms\cite{disha2022performance}. For instance, support vector machines (SVM) \cite{chen2005application} and Hidden Naïve Bayes (HNB) \cite{koc2012network} were proposed to enhance the accuracy and speed of the detection capability. The primary goal of ML-based IDS is to increase the number of correct predictions \cite{disha2022performance}, including the not-yet-known attacks (Zero-day attacks), which makes it more efficient than signature-based methods.  ML model quality can be evaluated using metrics, notably accuracy and F1 score \cite{disha2022performance}. Three main technical obstacles stall the building of ML-based IDS on the edge: (1) The considerable size of such a system renders implementations at the edge level a technical challenge, (2) required inference throughput on the resource-limited Edge-node hardware, and (3) update of the ML-based IDS requiring to re-initiate the whole system. Several approaches have been proposed to overcome these challenges, especially for the Edge-based deployment scenario. Therefore, a need exists for a clear methodology and criteria to build an ML-based IDS relying on reconfigurable Edge hardware.              

\subsection{Paper Contribution}
In this paper, we present a systematic selection methodology to construct a machine learning-based intrusion detection system targeting recon\-fi\-gur\-able Edge hardware. In particular, we first investigate the pros and cons of the reconfigurable Edge hardware in Section~\ref{sec:pors}. Two hardware configurations are selected: an FPGA-based dataflow processor and a RISC-V soft-core as an FPGA-based soft-core processor. Further, we establish hardware/software performance evaluation criteria for ML-based IDS (Intelligent IDS) on the Edge in Section~\ref{sec:criteria}. Then, we construct several machine-learning models to serve as an Intelligent IDS in Section~\ref{sec:intelligent_IDS}. Finally, we validate the established criteria against the detection systems running on the two proposed hardware configurations and evaluate their performance results in Section~\ref{Sec:Implementation}. Our approach aims at constructing an intelligent IDS relying on reconfigurable Edge hardware and providing high inference throughput to serve in high-performance Edge applications such as the future generation of high-speed communication technology.

To the best of our knowledge, this is the first work that establishes a systematic methodology for selecting a highly accurate ML-based IDS realized on two different configurations of reconfigurable Edge hardware. \section{Related Work}
In the following, we highlight the main approaches that leverage the reconfigurable hardware to build an ML-based IDS. 

\subsection{FPGA-based Dataflow Processor for ML-based IDS}
While the demand for FPGA-based dataflow processors (FPGA-based DFP) to accelerate ML and DNN algorithms using FPGAs grows, research on IDS designs in this area remains limited. FPGA-based DFP for ML-based IDS has been proposed in a few works, such as \cite{10.1007/978-3-030-34365-1_5} and \cite{9221584}. For example, a multilayer perceptron (MLP) was implemented on a Xilinx Virtex-5 FPGA in \cite{10.1007/978-3-030-34365-1_5}. The proposed network was trained on a smaller model with only six features from the NSL-KDD dataset \cite{8rpgqt9822}. It consists of two hidden layers. This MLP achieved a maximum throughput of 9.86\,Gbps with packets containing 1500 bytes featuring a speedup of 11.6$\times$ compared to a GPU. In \cite{9221584}, LogicNets \emph{``a methodology that allows trained quantized networks to be directly converted to an equivalent hardware''} \cite{9221584} was deployed to map a quantized MLP to hardware building blocks. The resulting DFP achieves a highly efficient acceleration rate. In \cite{10.1007/978-3-030-81645-2_9}, a convolutional neural network (CNN) topology on a PYNQ-Z2 was implemented. A quantization technique to explore 8-, 4-, and 2-bit quantization was employed. Extra pre-processing steps were also applied to reshape the raw data as an image. The experimentation used the CICIDS2017 dataset to detect one of 13 possible attack categories. The demonstrated DFP achieved a throughput of 9635 inferences/s at 100\,MHz with 99.4\% accuracy for the 2-bit quantized design. 

\subsection{FPGA-based Soft-Core Processor for ML-based IDS}
The use of RISC-V in intrusion detection and IoT security is the subject of recent research. A RISC-V SoC was proposed in \cite{10071321} as a platform to build a test environment for a man-in-the-middle attack simulation. In \cite{10076248}, a new RISC-V SoC was built based on the previous RISC-V SoC \cite{10071321} to construct a rule-based intrusion detection engine. The system runs Linux and uses Snort \cite{20031183} to capture network packets. If a match with the rules is found, an alarm will be triggered, and the event will be written into a log file. 

Several RISC-V soft cores to be used as a SCP have been proposed for performance-demanding and accelerated applications on the edge \cite{9885953,10.1145/1450095.1450107,10.1145/3468081.3471061}. In \cite{10139718}, an SCP (RISC-V CV32E41P) was synthesized to run at around 65\,MHz. The core is coupled to an on-FPGA tracer and arbiter to build a host-based IDS\cite{AZAD2008487}. Moreover, a different IDS implementation \cite{10139718} traces the hardware performance counters of the processor event values to detect any buffer overflow in the stack or heap in the Long-Range Wide Area Network (LoRaWan) protocol stack. 

Following the state-of-the-art, we developed our own approach towards FPGA-based DFP and SCP for use in ML-based IDS targeting reconfigurable edge hardware. We hence will first discuss the advantages of reconfigurable hardware for IDS on the edge in the following section. \section{Reconfigurable Edge Hardware for IDS: Pros \& Cons}
\label{sec:pors}
Besides the relatively lower cost of hardware design deployment on reconfigurable hardware (RHW) compared to other technologies, RHW enables tuning the hardware to current application needs, offering flexible update and extension of an implemented design. This feature also reduces development and re-engineering costs. FPGAs exhibit several advantages, such as low cost in the silicon chip area, high performance, and low power consumption \cite{kuon2006measuring,babu2021reconfigurable}. However, there are several limitations of RHW, most notably temporal or operational granularity. In the following, we highlight the pros and cons of FPGA-based DFP and SCP.  

\textbf{FPGA-based DFP} feature both a high level of parallelism and a need for reconfigurability. Their design offers high performance and low energy consumption, as highlighted by benchmarking studies \cite{10.1145/3373087.3375348}, particularly in DNN acceleration, making favorable comparisons with CPU and GPU platforms. However, the reconfigurability of FPGAs, while advantageous for computational acceleration, presents challenges due to the time- and power-consuming nature of the reconfiguration process. A trade-off has to be made between the static (running) phase and reconfiguration phases.
Despite the long-standing proposal of reconfigurable computing architecture, it has not gained widespread popularity. One reason for this is the requirement to use hardware design languages and dedicated design environments adding complexity and costs for developers \cite{8855594}. This is, however, mitigated by being able to perform a complete parallelization, hence allowing true parallel execution of operations without sacrificing inference accuracy. Parallelization and reducing an IDS's computational complexity are hence, the most prominent techniques used in an FPGA-based DFP. 

\textbf{FPGA-based SCP} being software-programmable by nature, are easier accessible by software programmers.They, however, also come at some cost to be considered when deciding to choose an IDS deployment platform. An FPGA-based SCP implementation offers:
\begin{itemize}
    \item \textbf{Flexibility} FPGA-based SCP can execute an IDS based on different computation precisions offered by the employed soft-core, such as Float32, INT8, or INT4. Orthogonally, soft-cores can be adjusted, enhanced, and extended, meeting new IDS requirements whenever needed.
    \item \textbf{Execution efficiency (performance)} With the availability of vector extensions to exploit data-parallel workloads \cite{9885953}, very efficient intrusion detection capability can be offered.
    \item \textbf{Portability} FPGA-based SCP can be implemented using cheaper FPGA resources, reducing overall system cost. Also, the code designed to run on a softcore might take advantage of high-level programming languages and libraries, thus making the developed code easily portable to other platforms. 
\end{itemize}
On the other hand, the development complexity and limited availability are just examples of some disadvantages of FPGA-based SCP. 
\begin{table}[!ht]
\centering
\caption{Comparison of FPGA-based DFP and SCP.}
\begin{tabular}{|c|c|c|c|}
\hline
\textbf{Reconfigurable} & \textbf{Computation} & \textbf{ML Topology} & \textbf{ML Parameters}\\
\textbf{Hardware} & \textbf{Precision} & \textbf{Update} & \textbf{Update}\\
\hline
\textit{FPGA-based DFP} & Fixed & Not On fly &  Partially \\
\hline
\textit{FPGA-based SCP} & Flexible & On fly & Flexible\\
\hline
\end{tabular}
\label{tab:ComDFPvsSCP}
\end{table}
 
Table~\ref{tab:ComDFPvsSCP} shows a comparison between FPGA-based DFP and SCP based on characteristics natively supported by the hardware, namely: computation precision, ML topology, ML parameter update, and required update time, that can deliver the best achievable performance and allow reconfigurability. 

FPGA-based DFP can easily accommodate ML hardware designs with floating point (FP), fixed point (FxP), and integer (INT) thanks to their reconfigurability. However, once an ML hardware design is programmed on the FPGA, it cannot be updated easily on the fly. Here, partial-dynamic reconfiguration could be a promising solution that allows a limited, predefined part of ML on an FPGA to be reconfigured while others continue working. Like any other CPU, FPGA-based SCP can easily accept any computation precision and update ML topology. In contrast, FPGA-based DFP requires repeating the process of generating a new hardware design to update ML topology.
The same goes for updating trained parameters: FPGA-based DFP require a particular mechanism for external parameter loading. This makes FPGA-based DFP partially able to update trained parameters. In the case of FPGA-based SCP, updating the ML topology or its trained parameters is comparatively less complicated, more straightforward, and less time-consuming. \section{Performance Evaluation Criteria for IDS on the Edge}
\label{sec:criteria}
To evaluate the performance of an IDS on the Edge, specific acceleration criteria must be considered on both levels, i.e. algorithm and reconfigurable hardware. We will detail this in the following two sections.
\subsection{IDS Algorithm Evaluation Criteria}
The IDS should be accurate from a software perspective, i.e., it should detect an intrusion with high accuracy and negligible false alarms. In the case of \textit{intelligent IDS}, several metrics can be used to evaluate how efficiently the ML model performs; these metrics can be highlighted as follows: 
\begin{itemize}
    \item \textbf{Precision (P)} This metric is fundamental as one goal of an \textit{intelligent IDS} is to minimize false positives. It measures how many of the positive predictions made are correct (true positives)\cite{muller2016introduction}.
    \item \textbf{Recall (R)} It measures how many positive cases the classifier correctly predicted over all the positive cases in the data. This metric is also important because an IDS aims to detect as many attacks as possible. \cite{doi:https://doi.org/10.1002/9781119556749.ch5}
    \item \textbf{F1 Score (F1)} described as the harmonic mean of the metrics \emph{Precision} and \emph{Recall} with both contributing equally to the score.\cite{muller2016introduction}.
\end{itemize}

Additionally, it should satisfy the following criterion: Even though there are several types of intrusion with different occurrence frequencies, an IDS should stay accurate when bias toward one attack over the other accrues. 

\subsection{Reconfigurable Hardware Evaluation Criteria}
In addition to meeting algorithmic requirements, also hardware criteria are to be met, which are:  
\begin{itemize}
    \item \textbf{Hardware Resource Utilization} An ideal IDS for edge devices should consume as few as possible resources, especially the DSP components, which are the most power-demanding units. This criteria significantly impacts the other hardware criteria, mainly computational density.   
    \item \textbf{Inference Throughput \cite{10.1145/3613963}} This criterion measures how many packets are processed by the intrusion detection system in a given amount of time. The IDS throughput is measured by $Packets/sec$. It should be noted that the network capacity limits this metric. \item \textbf{Energy Efficiency \cite{10.1145/3613963}} This criterion can be expressed as the inference throughput over energy consumption. For instance, the energy efficiency of the ML model for an IDS is evaluated by $Packets/sec/Watt$.\item \textbf{Computational Density \cite{10.1145/3613963}} Computational density is a metric used in FPGA design, referring to the ratio of computations performed by a particular design over the number of resources utilized. In other words, this criterion indicates whether the hardware design suffers from resource underutilization or not. For instance, when two different accelerators deliver the same inference throughput, the one with the lower DSP usage is considered better regarding computational density.
The computational density is expressed as $Throughput / \# DSP$ or $Throughput / \# LUT$.
    \item \textbf{Flexibility} used to measure and compare the complexity of development, maintenance, and new features implementation as well as maintainability and adaptability to new ML models and to new network conditions.
\end{itemize}

Some of these criteria directly impact the others. For instance, the computational density is directly derived from throughput and resource utilization. Likewise, resource utilization may indirectly decrease energy efficiency if the resources are too power-demanding compared to the achieved throughput. \section{Proposed IDS Design Methodology}
\label{sec:intelligent_IDS}
The previous discussion is applied to a 4-step approach in order to design, implement, and evaluate FPGA-based DFP and SCP approaches. We 1) construct several IDSs based on state-of-the-art algorithms in the ML domain, then 2) select ML models with high precision (P) and F1 scores and smaller model sizes in terms of byte, before 3) implementing the chosen ML for IDS on FPGA-based DFP and SCP and finally 4) analyzing these implementations regarding the proposed hardware criteria and, following a dedicated edge use-case, choosing the IDS implementation that matches the high-speed requirements of modern communication technology.

\subsection{Step 1: Intelligent IDS}
\label{sectelligent_IDS}
In this section, we outline the use and customization of several well-known ML models and MLP, on which our IDS is based. Choosing and adjusting the right model is paramount for both, detection quality and hardware use. As shown in the evaluation section, the resulting \textit{Intelligent IDS} can offer vast performance at minimal hardware cost with NN capability at discerning intricate patterns in extensive datasets. 

\subsubsection{The BOT-IoT Dataset}
\begin{figure}[ht]
    \centering
    \includegraphics[width=0.8\linewidth]{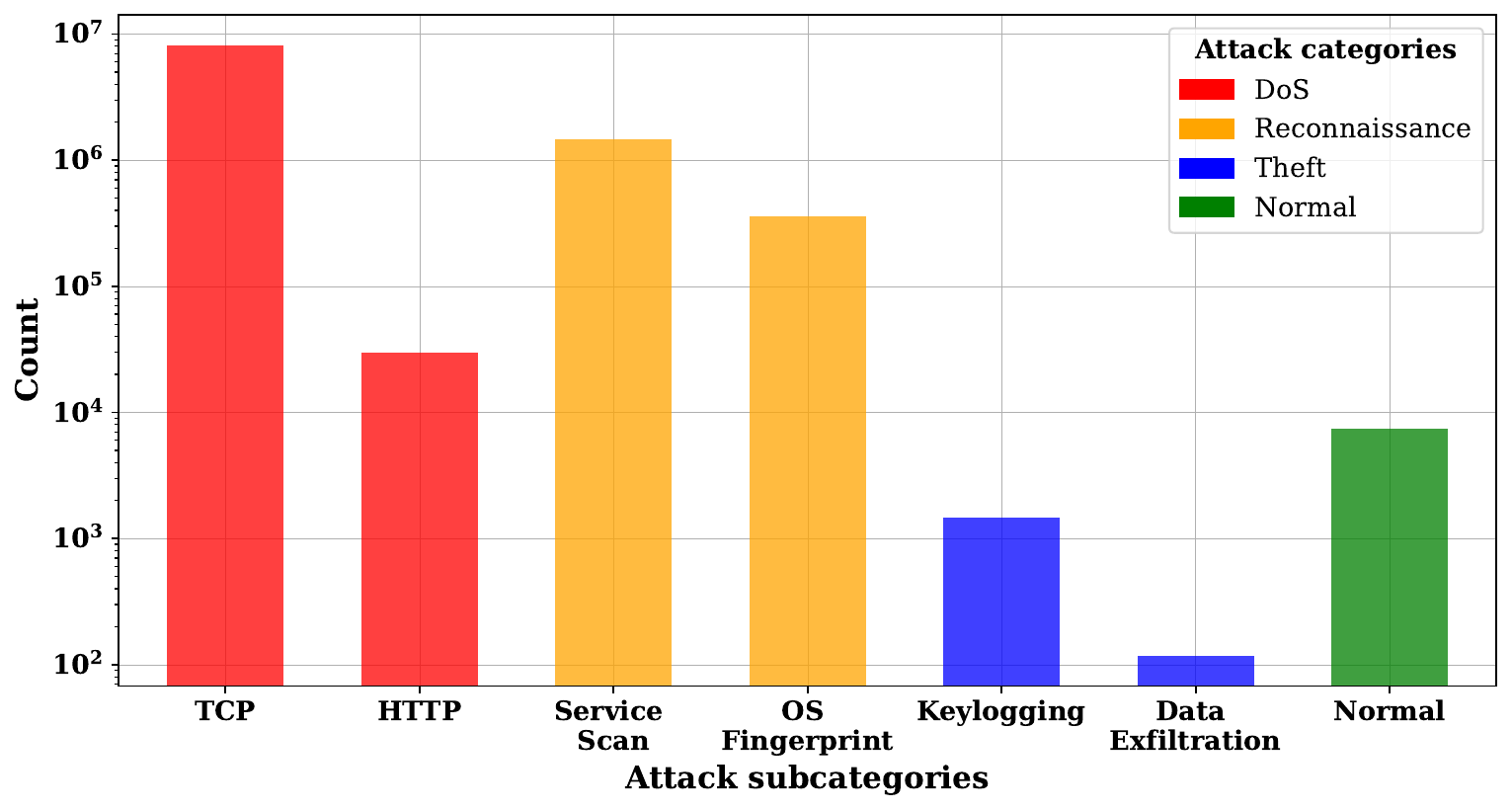}
    \caption{Attack categories and subcategories distribution of the BOT-IoT dataset}
    \label{fig:data_bal}
\end{figure}

Bot-IoT \cite{koroniotis2019towards} was developed within a testbed environment, employing a constellation of virtual machines featuring diverse operating systems, network firewalls, network taps, the Node-red, and the Argus tools \cite{node2021node,argus}. The Bot-IoT dataset is characterized by multiple sets and subsets, each distinguished by file format, size, and feature count variations. Fig~\ref{fig:data_bal} shows the dataset balance for each attack category and subcategory, reflecting the whole dataset's general imbalance. The BOT-IoT dataset includes several attack scenarios. From these, we select a subset that covers the following attacks: DoS (TCP, HTTP), reconnaissance (service scan and OS fingerprinting), theft (keylogging and data extraction), and intrusion-free.        

\subsubsection{Intelligent IDS Construction}
This step involves partitioning the pre-processed data into an 80\% training set and a 20\% set for testing and evaluation. We first start with training XGBoost (XGB), Support Vector Machine (SVM), Naive Bayes (NB), Random Forest Classifier (RFC), and Decision Tree (DT). Additionally, three Multi-Layer Perceptron (MLP) models are trained, each of them tailored to distinct classification targets: attacks, categories, and subcategories. All of the three MLP models share a nearly identical topology, featuring an input layer with 24 inputs, followed by two hidden layers of sizes 32 and 64, respectively, and Rectified Linear Unit (\emph{ReLU}) activation functions. The sole distinction among the MLP models resides in the configuration of the final layer, i.e., the classification layer. The \emph{Attack} model is designed to discern the presence or absence of an attack; hence, its last layer has a size of 2 followed by a \emph{Softmax} activation. Analogously, the \emph{Category} and \emph{Subcategory} models' classification layers exhibit sizes of 4 and 7, respectively, aligning with their distinct classification objectives. 

\subsection{Step 2: Intelligent IDS Selection}
According to the introduced algorithm evaluation criteria, a performance comparison for each ML model is made. Table~\ref{tab:ML_Alg_comp} compares model detection accuracy and size. NB exhibits a very low F1 Score. Therefore, it will be eliminated and we focus on ML models with a high F1 Score. Overall, XGboost and MLP outperform other models. 
\begin{table}[!ht]
\centering
\caption{Evaluation of ML algorithms for IDS: ML Metrics vs Size.}
\begin{tabular}{|c||c|c|c|c||c|c|c|c||c|c|c|c|}
\hline
\multirow{3}{*}{\textbf{Algorithm}} & \multicolumn{4}{c||}{\textbf{Attack}} & \multicolumn{4}{c||}{\textbf{Category}} & \multicolumn{4}{c|}{\textbf{Subcategory}} \\ 
                                    & \multicolumn{4}{c||}{\textbf{detection}} & \multicolumn{4}{c||}{\textbf{classification}} & \multicolumn{4}{c|}{\textbf{classification}} \\
\cline{2-13}
            & $P$ & $R$ & $F_1$ & size & $P$ & $R$ & $F_1$ & size & $P$ & $R$ & $F_1$ & size \\ \hline
\textbf{XGB} & 1.00 & 1.00 & 1.00 & 0.38 MB & 1.00 & 1.00 & 1.00 & 1.15 MB & 0.99 & 0.99 & 0.99 & 2.15 MB\\ 
\hline
\textbf{SVM} & 0.99 & 0.99 & 0.99 & 164 KB & 1.00 & 0.99 & 1.00 & 288 KB & 0.97 & 0.89 & 0.92 & 5.7 MB \\ \hline
\textbf{NB} & 0.57 & 0.96 & 0.62 & 1.35 KB & 0.78 & 0.94 & 0.78 & 2.23 KB & 0.78 & 0.70 & 0.60 & 3.62 KB\\ 
\hline
\textbf{RFC} & 1.00 & 0.97 & 0.98 & 123 KB & 1.00 & 0.95 & 0.97 & 157 KB & 0.70 & 0.59 & 0.62 & 0.2 MB\\ 
\hline
\textbf{DT} & 0.99 & 0.99 & 0.99 & 2.23 KB & 1.00 & 0.99 & 1.00 & 3.25 KB & 0.83 & 0.81 & 0.82 & 4.35 KB\\ 
\hline
\textbf{MLP} & 1.00 & 1.00 & 1.00 & 15.2 KB & 1.00 & 1.00 & 1.00 & 15.8 KB & 0.99 & 0.93 & 0.96 & 16.6 KB\\ 
\hline
\end{tabular}
\label{tab:ML_Alg_comp}
\end{table}
 
\subsection{Step 3: FPGA-based Intelligent IDS Implementation}
\label{Sec:Implementation}
Here, we describe the experimental setup of the FPGA-based DFP and the RISC-V soft-core as FPGA-based SCP. Both experiments are evaluated using the Xilinx ZCU104 FPGA platform.

\subsubsection{Experimental Setup}
\begin{figure}
    \centering
    \includegraphics[width=0.925\linewidth]{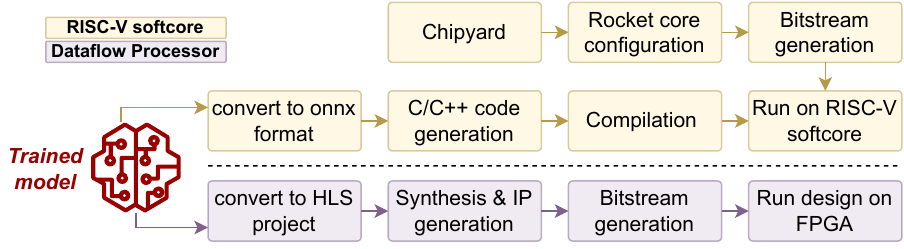}
    \caption{IDS Experimental Setup RISC-V Soft-core and the FPGA-based DFP.}
    \label{fig:expr}
\end{figure}

The experimental setup, illustrated in Fig.~\ref{fig:expr}, includes the FPGA-based RISC-V SCP and the DFP experimental process. Opting for a 64-bit Rocket core \cite{Asanovic:EECS-2016-17}, configured through the Chipyard framework \cite{chipyard}, the Rocket core stands as a 5-stage single-issue in-order processor executing the 64-bit scalar RISC-V ISA\cite{Waterman:EECS-2016-1}. This core can accommodate operating systems and features an optional IEEE\,754-2008-compliant FPU for single- and double-precision floating-point operations, including fused multiply-accumulate. MLP models are saved as ONNX models, transformed into C code for seamless porting onto a RISC-V soft-core, and compiled using the appropriate RISC-V GNU toolchain and flags\footnote{\texttt{\$ riscv64-unknown-elf-gcc -std=gnu99 -O2 -Wall -lm -fno-common -fno-builtin-printf -specs=htifnano.specs} \\* \texttt{\$ riscv64-unknown-elf-gcc -static -T riscv64-unknown-elf/lib/htif.ld -lm}} for bare-metal execution.

In contrast, the approach for the FPGA-based DFP involved converting trained models to HLS projects using hls4ml. Notably, hls4ml lacked inherent support for Float32 conversion, prompting manual intervention to adjust data types for different layers and activation functions. The Softmax function is also modified to accommodate Float32 operations, ensuring a fair comparison between the FPGA-based DFP and SCP. Subsequently, IPs for various MLP models are generated and integrated into a corresponding FPGA design, and the resulting Bitstreams are deployed on the FPGA platform for benchmarking.

\subsection{Step 4: Systematic Implementation Comparison}
\subsubsection{Hardware Usage Comparison:} 
\begin{figure}[!ht]
    \centering
    \includegraphics[width=0.9\linewidth]{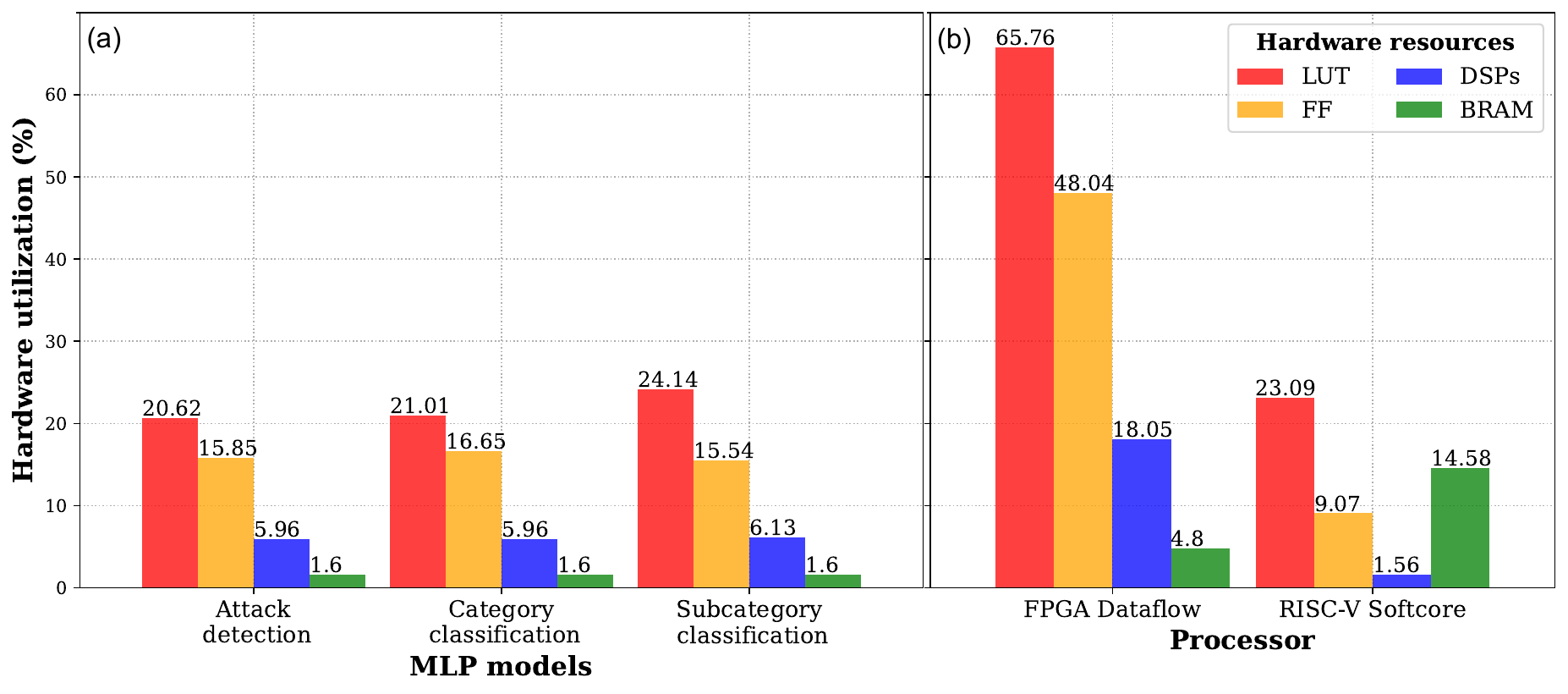}
    \caption{Hardware Utilization of (a) our 3 MLPs on FPGA and (b) FPGA-based RISC-V SCP vs. Overall MLPs as FPGA-based DFP.}
    \label{fig:utils_dataflow}
\end{figure}

Fig.~\ref{fig:utils_dataflow}-(a) shows the required hardware resources to implement FPGA-based DFP as three individual MLP IPs. All the IPs exhibit identical Block RAM (BRAM) utilization ratios for FPGA-based DFP. This uniformity can be attributed to the shared topology among MLP models, except for the last layer. The shared structure comprises two layers, constituting the most memory-intensive part due to their incorporation of most model parameters. However, slight variations in other resources arise from the intentional partitioning of parameters and result arrays in the last Softmax layer, mapped as Look-Up Tables (LUTs) and First-In-First-Out (FIFO) structures. The size of the Softmax layer accounts for the marginal fluctuations in the use of the Digital Signal Processors (DSP).

To compare FPGA-based DFP and RISC-V SCP, we implement them to operate at the same frequency of 100\,MHz. Their respective hardware utilization ratios are illustrated in Fig.~\ref{fig:utils_dataflow}-(b). The high parallelization of the FPGA-based DFPs requires more resources than the FPGA-based RISC-V SCP, except for the BRAM units, which seem to be used more by the softcore for the caches. In contrast, FPGA-based DFPs require more LUTs, FIFOs, and DSPs for parallelized processing.

\subsubsection{Comparison of Throughput, Energy Efficiency \& Logic density}
\label{sec:perf_comparaison}
We analyze and compare the two designs based on the earlier-defined criteria. The FPGA-based DFP is configured so that every compute unit executes only four multiply-accumulate operations sequentially, resulting in a higher parallelism. A design parameter, namely \emph{Reuse factor}, controls such a parallelism mechanism. Additionally, the last Softmax layer was fully unrolled to compensate for the extra latency overhead caused by using Float32 arithmetic. As a result, FPGA-based DFP exhibits $\approx 6\times$ higher throughput than the FPGA-based RISC-V SCP, as shown in Table~\ref{tab:res_ac}. FPGA-based RISC-V SCP, in the term, draws only 2.34\,W, almost half the power of the FPGA-based DFP. However, its throughput superiority makes the latter $\approx 3\times$ more energy efficient than FPGA-based RISC-V SCP. Also, this is why it exhibits between 5 and 6x higher logic density. These measures can undoubtedly be even higher with low-precision arithmetic such as fixed-point and integer, especially for FPGA-based DFP.
\begin{table}[!ht]
\centering
\caption{FPGA-based DFP vs RISC-V SCP with Float32 Precision}
\scalebox{0.9}{
\begin{tabular}{|c|c|c|c|}
\hline
\multirow{2}{*}{\textbf{MLP Model}} & \textbf{Throughput} & \textbf{Energy Efficiency} & \textbf{Logic Density} \\&  $Packets/sec$      & $Packets/sec/W$  & $Packets/sec/LUT$ \\
\hline
\multicolumn{4}{|c|}{\textbf{FPGA-based DFP}}\\
\hline
\textit{Attack} & 1166861 (1.16 M) & 265799 (265 K)& 24.55 \\\hline
\textit{Category} & 1135073 (1.13 M) & 255589 (255 K)& 23.44 \\\hline
\textit{Subcategory} & 1118568 (1.11 M) & 249346 (249 K) & 20.11 \\\hline
\multicolumn{4}{|c|}{\textbf{RISC-V SCP - Optimized Baremetal}}\\
\hline
\textit{Attack} & 202849 (202 K)& 86650 (86 K)& 4.157 \\\hline
\textit{Category} & 197500 (197 K)& 84365 (84 K)& 4.047 \\\hline
\textit{Subcategory} & 197342 (197 K)& 84298 (84 K)& 4.44 \\\hline
\end{tabular}}
\label{tab:res_ac}
\end{table}
 
\subsubsection{Flexibility Comparison}
Both processing systems have been evaluated based on their flexibility as detailed in Table~\ref{tab:flex}. The flexibility comparison is dedicated to the implemented processors and is based on the above-mentioned aspects: precision, topology, and parameter updates. Additionally, we investigate the required time to update.
\begin{table}[!ht]
\centering
\caption{Flexibility Comparison of FPGA-based DFP and RISC-V SCP.}
\begin{threeparttable}
\scalebox{0.90}{
\begin{tabular}{|c|c|c|c|c|c|c|}
\hline
\multirow{2}{*}{\textbf{Processor}} & \multicolumn{3}{c|}{\textbf{Precision}} & \textbf{Topology} & \textbf{Parameters} & \textbf{Update}\\
\cline{2-4}
                   & \textbf{FP} & \textbf{FxP} & \textbf{INT}  &               \textbf{update}                     &         \textbf{update}           & \textbf{time} \\
\hline
\textit{FPGA-based DFP} & yes & yes & yes & no & no\tnote{1} & longer\\
\hline
\textit{FPGA-based RISC-V SCP} & yes & no & yes\tnote{2} & yes & yes & shorter\\
\hline
\end{tabular}}
\begin{tablenotes}\scriptsize
\item[1] Only possible if the design includes an external weights loading mechanism. 
\item[2] Supports only a subset of integers, such as INT8/16/32.
\end{tablenotes}
\end{threeparttable}
\label{tab:flex}
\end{table} The proposed FPGA-based DFP is very flexible regarding computation precision, such as floating point (FP), fixed point (FxP), and integer (INT) due to FPGA reconfigurability. The chosen FPGA-based RISC-V SCP, in term, has a fixed data-path, which limits its precision capability to FP and a specific set of integers, such as INT8, 16, and 32. Consequently, it offers fewer options to optimize the ML-based IDS through quantization. However, updating the ML topology or its trained parameters is significantly less complicated and, therefore, more straightforward in the case of FPGA-based RISC-V SCP; it only requires a new source-code compilation.

\subsubsection{Proposed FPGA-based DFP Compared to the State of the Art}  
\begin{table}[h]
\centering
\caption{State of the Art FPGA-based DFP for ML-based IDS.}

\scalebox{0.85}{
\begin{tabular}{|c|c|c|c|c|c|c|c|}
\hline
\textbf{References} & \textbf{\cite{10.1007/978-3-030-81645-2_9}} & \textbf{\cite{9221584}} & \textbf{\cite{10.1007/978-3-030-34365-1_5}} & \textbf{\cite{ioannou2019network}} & \multicolumn{3}{c|}{\textbf{This work}}\\
\hline
\textit{FPGA}       & xc7Z020 & xc7Z020    & xc5vtx & xc7Z020  & \multicolumn{3}{c|}{xczu7ev} \\
\hline
\textit{Frequency}  & \multirow{2}{*}{100}   & \multirow{2}{*}{471} & \multirow{2}{*}{104} & \multirow{2}{*}{76} & \multicolumn{3}{c|}{\multirow{2}{*}{100}}\\
\textit{(MHz)}  & & & & & \multicolumn{3}{c|}{ }\\
\hline
\textit{Dataset}    & CICIDS2017\cite{sharafaldin2018a}  & UNSW-NB15\cite{7348942}      & NSL-KDD\cite{8rpgqt9822} & NSL-KDD  & \multicolumn{3}{c|}{BOT-IoT\cite{koroniotis2019towards}} \\
\hline
\textit{ML topology} & CNN & MLP & MLP & MLP & \multicolumn{3}{c|}{MLP}  \\
\hline
\textit{Number} & 4$\times$Conv + & \multirow{2}{*}{5$\times$FC} & \multirow{2}{*}{2$\times$FC} & \multirow{2}{*}{3$\times$FC} & \multicolumn{3}{c|}{\multirow{2}{*}{3$\times$FC}}\\
\textit{of layers} & 2$\times$FC &  &  &  &  \multicolumn{3}{c|}{}\\
\hline
\textit{Intrusion Classes}  & 13 & 2 & 2 & 2 & 2 & 4 & 7\\
\hline
\textit{Accuracy (\%)} & 99.4 & 91.3 & 87.3  & 80.52 & 99.9 & 99.9 & 99.9 \\
\hline
\textit{Throughput} & \multirow{2}{*}{9635} & \multirow{2}{*}{754292} & \multirow{2}{*}{821667} & \multirow{2}{*}{217074} & 1.16 & 1.13 & 1.11\\
\textit{(Packets/sec)} &  &  &  &  & M & M & M\\
\hline
\textit{LUT usage}  & 24635 & 15494 &  117082  & 26463  & 47514  & 48413 & 55627 \\   
\hline
\textit{Usage ratio (\%)}  & 46.3  &  29.12 & 78.2 & 50 &  20.6  & 21 & 24.1\\
\hline
\end{tabular}}
\label{tab:SoACom}
\end{table} 
Table~\ref{tab:SoACom} compares our proposed FPGA-based DFP for ML-based IDS and the state of the art in this domain. The results show our proposed intelligent IDS detects 13 different intrusion classes, and its implementation as FPGA-based DFP exhibits very high throughput, yet low hardware resources. This makes it suitable for application at the edge level and clearly demonstrates that such a high-performance solution does not necessarily come at prohibitively high hardware costs.
 \section{Conclusion}
Current modern approaches to intrusion detection systems (IDS) feature the use of machine learning (ML). However, ML-based IDSs still face technical obstacles such as their considerable size, and their update requires re-initiating the whole IDS. In this paper, we investigate ML-based IDS targeting the edge level, featuring reconfigurable edge nodes. 
Here, typically high throughput is required in order to keep up with the real-time data transmissions, yet node resource use is constrained. Orthogonally, intrusion detection in a reconfigurable system also demands an equally flexible adaptability with respect to detection itself.
We hence construct a systematic approach to ML-based intrusion detection on the edge, leading to the proposed Intelligent IDS. We discuss two possible FPGA-based implementation scenarios, one plain hardware implementation (FPGA-based dataflow processor, DFP) and one featuring a RISC-V softcore.
Both implementations are evaluated and compared to each other and the state of the art. The results clearly demonstrate that the high performance of a hardware implementation does not necessarily come at prohibitively high hardware cost, with our solution
exhibiting higher throughput, better energy efficiency, and better logic density in addition to an overall better configurability. Our proposed DFP hence can be employed in high-performance Edge-based applications like modern communication technology.
 
\section*{Disclosure of Interests}
The authors have no competing interests to declare that are relevant to the content of this article.

\bibliographystyle{splncs04}

\begin{thebibliography}{10}
\providecommand{\url}[1]{\texttt{#1}}
\providecommand{\urlprefix}{URL }
\providecommand{\doi}[1]{https://doi.org/#1}

\bibitem{chipyard}
Amid, A., Biancolin, D., Gonzalez, A., Grubb, D., Karandikar, S., Liew, H.,
  Magyar, A., Mao, H., Ou, A., Pemberton, N., Rigge, P., Schmidt, C., Wright,
  J., Zhao, J., Shao, Y.S., Asanovi\'{c}, K., Nikoli\'{c}, B.: Chipyard:
  Integrated design, simulation, and implementation framework for custom socs.
  IEEE Micro  \textbf{40}(4),  10--21 (2020). \doi{10.1109/MM.2020.2996616}

\bibitem{20031183}
Amon, C., Shinder, T.W., Carasik-Henmi, A.: Chapter 29 - introducing snort. In:
  The Best Damn Firewall Book Period, pp. 1183--1208. Syngress, Burlington
  (2003). \doi{10.1016/B978-193183690-6/50070-4}

\bibitem{Asanovic:EECS-2016-17}
Asanović, K., Avizienis, R., Bachrach, J., Beamer, S., Biancolin, D., Celio,
  C., Cook, H., Dabbelt, D., Hauser, J., Izraelevitz, A., Karandikar, S.,
  Keller, B., Kim, D., Koenig, J., Lee, Y., Love, E., Maas, M., Magyar, A.,
  Mao, H., Moreto, M., Ou, A., Patterson, D.A., Richards, B., Schmidt, C.,
  Twigg, S., Vo, H., Waterman, A.: The rocket chip generator. Tech. Rep.
  UCB/EECS-2016-17, EECS Department, University of California, Berkeley (Apr
  2016)

\bibitem{AZAD2008487}
Azad, T.B.: Chapter 7 - locking down your xenapp server. In: Azad, T.B. (ed.)
  Securing Citrix Presentation Server in the Enterprise, pp. 487--555.
  Syngress, Burlington (2008). \doi{10.1016/B978-1-59749-281-2.00007-X}

\bibitem{babu2021reconfigurable}
Babu, P., Parthasarathy, E.: Reconfigurable fpga architectures: A survey and
  applications. Journal of The Institution of Engineers: Series B
  \textbf{102},  143--156 (2021)

\bibitem{10.1145/3373087.3375348}
Blott, M., Kath, J., Halder, L., Umuroglu, Y., Fraser, N., Gambardella, G.,
  Leeser, M., Doyle, L.: Evaluation of optimized cnns on fpga and non-fpga
  based accelerators using a novel benchmarking approach. In: Proceedings of
  the 2020 ACM/SIGDA International Symposium on Field-Programmable Gate Arrays.
  p.~317. FPGA '20, Association for Computing Machinery, New York, NY, USA
  (2020). \doi{10.1145/3373087.3375348}

\bibitem{10139718}
Bouazzati, M.E., Tessier, R., Tanguy, P., Gogniat, G.: A lightweight intrusion
  detection system against iot memory corruption attacks. In: 2023 26th
  International Symposium on Design and Diagnostics of Electronic Circuits and
  Systems (DDECS). pp. 118--123 (2023). \doi{10.1109/DDECS57882.2023.10139718}

\bibitem{Forbes2016}
Brewster, T.: How hacked cameras are helping launch the biggest attacks the
  internet has ever seen. Forbes (2016),
  https://www.forbes.com/sites/thomasbrewster/2016/09/25/brian-krebs-overwatch-ovh-smashed-by-largest-ddos-attacks-ever/

\bibitem{10071321}
Cai, B., Xie, S., Liang, Q., Lu, W.: Research on penetration testing of iot
  gateway based on risc- v. In: 2022 International Symposium on Advances in
  Informatics, Electronics and Education (ISAIEE). pp. 422--425 (2022).
  \doi{10.1109/ISAIEE57420.2022.00093}

\bibitem{9885953}
Chander, V.N., Varghese, K.: A soft risc-v vector processor for edge-ai. In:
  2022 35th International Conference on VLSI Design and 2022 21st International
  Conference on Embedded Systems (VLSID). pp. 263--268 (2022).
  \doi{10.1109/VLSID2022.2022.00058}

\bibitem{chen2005application}
Chen, W.H., Hsu, S.H., Shen, H.P.: Application of svm and ann for intrusion
  detection. Computers \& Operations Research  \textbf{32}(10),  2617--2634
  (2005)

\bibitem{denning1987intrusion}
Denning, D.E.: An intrusion-detection model. IEEE Transactions on software
  engineering  \textbf{SE-13}(2),  222--232 (1987)

\bibitem{disha2022performance}
Disha, R.A., Waheed, S.: Performance analysis of machine learning models for
  intrusion detection system using gini impurity-based weighted random forest
  (giwrf) feature selection technique. Cybersecurity  \textbf{5}(1), ~1 (2022)

\bibitem{1106666}
Hutchings, B., Franklin, R., Carver, D.: Assisting network intrusion detection
  with reconfigurable hardware. In: Proceedings. 10th Annual IEEE Symposium on
  Field-Programmable Custom Computing Machines. pp. 111--120 (2002).
  \doi{10.1109/FPGA.2002.1106666}

\bibitem{ioannou2019network}
Ioannou, L., Fahmy, S.A.: Network intrusion detection using neural networks on
  fpga socs. In: 2019 29th International Conference on Field Programmable Logic
  and Applications (FPL). pp. 232--238. IEEE (2019)

\bibitem{10.1145/3468081.3471061}
Kimura, Y., Ootsu, K., Tsuchiya, T., Yokota, T.: Development of risc-v based
  soft-core processor with scalable vector extension for embedded system. In:
  Proceedings of the the 8th International Virtual Conference on Applied
  Computing \& Information Technology. p. 13–18. ACIT '21, Association for
  Computing Machinery, New York, NY, USA (2021). \doi{10.1145/3468081.3471061}

\bibitem{koc2012network}
Koc, L., Mazzuchi, T.A., Sarkani, S.: A network intrusion detection system
  based on a hidden na{\"\i}ve bayes multiclass classifier. Expert Systems with
  Applications  \textbf{39}(18),  13492--13500 (2012)

\bibitem{koroniotis2019towards}
Koroniotis, N., Moustafa, N., Sitnikova, E., Turnbull, B.: Towards the
  development of realistic botnet dataset in the internet of things for network
  forensic analytics: Bot-iot dataset. Future Generation Computer Systems
  \textbf{100},  779--796 (2019)

\bibitem{kuon2006measuring}
Kuon, I., Rose, J.: Measuring the gap between fpgas and asics. In: Proceedings
  of the 2006 ACM/SIGDA 14th international symposium on Field programmable gate
  arrays. pp. 21--30 (2006)

\bibitem{10.1007/978-3-030-81645-2_9}
Le~Jeune, L., Goedem{\'e}, T., Mentens, N.: Towards real-time deep
  learning-based network intrusion detection on fpga. In: Applied Cryptography
  and Network Security Workshops. pp. 133--150. Springer International
  Publishing, Cham (2021)

\bibitem{10076248}
Liang, Q., Xie, S., Cai, B.: Intelligent home iot intrusion detection system
  based on risc-v. In: 2023 IEEE 3rd International Conference on Power,
  Electronics and Computer Applications (ICPECA). pp. 296--300 (2023).
  \doi{10.1109/ICPECA56706.2023.10076248}

\bibitem{doi:https://doi.org/10.1002/9781119556749.ch5}
Mishra, A.: Evaluating Machine Learning Models, chap.~5, pp. 115--132. John
  Wiley and Sons, Ltd (2019). \doi{10.1002/9781119556749.ch5}

\bibitem{7348942}
Moustafa, N., Slay, J.: Unsw-nb15: a comprehensive data set for network
  intrusion detection systems (unsw-nb15 network data set). In: 2015 Military
  Communications and Information Systems Conference (MilCIS). pp.~1--6 (2015).
  \doi{10.1109/MilCIS.2015.7348942}

\bibitem{muller2016introduction}
M{\"u}ller, A.C., Guido, S.: Introduction to machine learning with Python: a
  guide for data scientists. O'Reilly Media, Inc. (2016)

\bibitem{10.1145/3613963}
Nechi, A., Groth, L., Mulhem, S., Merchant, F., Buchty, R., Berekovic, M.:
  Fpga-based deep learning inference accelerators: Where are we standing? ACM
  Trans. Reconfigurable Technol. Syst.  \textbf{16}(4) (oct 2023).
  \doi{10.1145/3613963}

\bibitem{10.1007/978-3-030-34365-1_5}
Ngo, D.M., Tran-Thanh, B., Dang, T., Tran, T., Thinh, T.N., Pham-Quoc, C.:
  High-throughput machine learning approaches for network attacks detection on
  fpga. In: Vinh, P.C., Rakib, A. (eds.) Context-Aware Systems and
  Applications, and Nature of Computation and Communication. pp. 47--60.
  Springer International Publishing, Cham (2019)

\bibitem{node2021node}
Node-RED: Low-code programming for event-driven applications (2021),
  \url{https://nodered.org/}

\bibitem{argus}
QOSIENT, L.: Argus (2023), \url{https://openargus.org/}

\bibitem{sha2020survey}
Sha, K., Yang, T.A., Wei, W., Davari, S.: A survey of edge computing-based
  designs for iot security. Digital Communications and Networks  \textbf{6}(2),
   195--202 (2020)

\bibitem{sharafaldin2018a}
Sharafaldin, I., Lashkari, A.H., Ghorbani, A.A.: Toward generating a new
  intrusion detection dataset and intrusion traffic characterization. In: 4th
  International Conference on Information Systems Security and Privacy (ICISSP.
  Portugal (2018)

\bibitem{9221584}
Umuroglu, Y., Akhauri, Y., Fraser, N.J., Blott, M.: Logicnets: Co-designed
  neural networks and circuits for extreme-throughput applications. In: 2020
  30th International Conference on Field-Programmable Logic and Applications
  (FPL). pp. 291--297 (2020). \doi{10.1109/FPL50879.2020.00055}

\bibitem{vasilomanolakis2015taxonomy}
Vasilomanolakis, E., Karuppayah, S., M{\"u}hlh{\"a}user, M., Fischer, M.:
  Taxonomy and survey of collaborative intrusion detection. ACM Computing
  Surveys (CSUR)  \textbf{47}(4),  1--33 (2015)

\bibitem{8855594}
Wang, T., Wang, C., Zhou, X., Chen, H.: An overview of fpga based deep learning
  accelerators: Challenges and opportunities. In: 2019 IEEE 21st International
  Conference on High Performance Computing and Communications; IEEE 17th
  International Conference on Smart City; IEEE 5th International Conference on
  Data Science and Systems (HPCC/SmartCity/DSS). pp. 1674--1681 (2019).
  \doi{10.1109/HPCC/SmartCity/DSS.2019.00229}

\bibitem{Waterman:EECS-2016-1}
Waterman, A.: Design of the RISC-V Instruction Set Architecture. Ph.D. thesis,
  EECS Department, University of California, Berkeley (Jan 2016),
  \url{http://www2.eecs.berkeley.edu/Pubs/TechRpts/2016/EECS-2016-1.html}

\bibitem{10.1145/1450095.1450107}
Yiannacouras, P., Steffan, J.G., Rose, J.: Vespa: Portable, scalable, and
  flexible fpga-based vector processors. In: Proceedings of the 2008
  International Conference on Compilers, Architectures and Synthesis for
  Embedded Systems. p. 61–70. CASES '08, Association for Computing Machinery,
  New York, NY, USA (2008). \doi{10.1145/1450095.1450107}

\bibitem{8rpgqt9822}
ZHAO, R.: Nsl-kdd (2022). \doi{10.21227/8rpg-qt98}

\end{thebibliography}

\end{document}